\documentclass[reprint,amsmath,amssymb,aps]{revtex4-1}

\usepackage{graphicx}
\usepackage{dcolumn}
\usepackage{bm}

\begin{document}

\preprint{APS/123-QED}

\title{Gaseous plasmonic resonators for metamaterial applications}

\author{Roberto A. Col\'on Qui\~nones}
\email{racolon@stanford.edu}
\author{Thomas C. Underwood}
\author{Mark A. Cappelli}

\affiliation{High Temperature Gasdynamics Laboratory, Stanford University, Stanford, CA 94305, USA}

\date{\today}

\begin{abstract}
We examine the properties of a gaseous plasma resonator generated by focusing a high-energy laser pulse through a lens and into a gas. An analytical model is presented describing the scattering resonance of these near-ellipsoidal plasmas and its dependence on their eccentricity and intrinsic plasma properties. This dependence is investigated through Ku band transmission experiments of a waveguide with an embedded single plasma element and through optical diagnostics of the laser-induced plasma. The described resonator has the potential to be used as the building block in a new class of metamaterials with fully three-dimensional structural flexibility.
\end{abstract}

\maketitle

The electromagnetic (EM) properties of most natural materials are dictated by the resonances of their constituent atoms, which are typically $\sim$ 1 angstrom in size. Because of the small spacing between these atoms, many EM wavelengths of engineering interest (vacuum ultraviolet waves or longer) only probe their average or effective properties. A metamaterial (MTM), i.e., an artificial material engineered to exhibit EM properties not readily found in nature, follows the same principle \cite{smith2004}. The resonantly active elements in  MTMs and their relative spacing are designed to create an effectively homogeneous medium with engineered EM properties. Such a medium is achieved when the spacing between the constituent elements is smaller than a quarter of the guided wavelength \cite{Caloz2005}. The ability to create MTMs with exotic effective properties has lead to the formation of the very active field of transformation optics where MTMs of spatially varying refractive indexes are used to steer EM waves in unusual ways to achieve phenomena such as negative refraction, perfect lensing, and optical cloaking \cite{chen2010transformation}.
 
In recent years, a significant portion of research in this field has focused on exploring the tunability of these phenomena. Tunable MTM devices have exploited the electromechanical, electro-optical, liquid-crystal, phase-change, and superconducting response of their constituent elements \cite{zheludev2012}. One class of potentially tunable elements that has been largely overlooked are gaseous plasmas, which can be incorporated into more traditional MTMs to form composite periodic structures with tunable features \cite{Sakai2012}. The tunability of these structures stems from the dispersive nature of plasmas arising from their variable electron density ($n_e$) and electron momentum transfer collision frequency ($\gamma$). These two plasma properties are in turn controlled by the energy invested by the ionizing source and the pressure/composition of the gas. 

The majority of past research related to plasma MTMs has focused mainly on the development of composites that integrate plasmas into metallic resonant structures \cite{kim2018spatially,kourtzanidis2016,Lee2009,lee2017plasma,Liu2015,nakamura2014,qu2017properties,Sakai2010,singh2014metamaterials} or into dielectric resonator arrays \cite{cohick2016,dennison2016plasma}. Waveguiding MTM structures composed exclusively of laser-induced plasma (LIP) filaments have also been proposed \cite{kudyshev2013virtual}, but have yet to be realized experimentally. This letter reports on a gaseous plasma resonator that could potentially be used as the building block for an all-plasma MTM. Specifically, this study describes the EM properties of a LIP resonator which is generated by tightly focusing the fundamental output from a high-power laser through a lens and into a gas at constant pressure \cite{demichelis1969,ostrovskaya1974laser}. The near-ellipsoidal \cite{chen2000spatial,roskos2007broadband}, sub-wavelength plasma resonator interacts with incoming radiation through excitation of low-order, electric-dipole resonances similar to those seen in metallic spheres \cite{bohren1998}. For frequencies below the plasma frequency ($\omega_p$), the collective scattering response of a MTM composed entirely of these plasmonic resonators results in a finite region of negative effective dielectric constant ($\epsilon$) near the frequency of the excited surface mode \cite{zhao2009mie}. Potential applications of such MTMs include, but are not limited to, highly-tunable reflective surfaces and negative-index mediums with fully three-dimensional structural flexibility. These MTMs are particularly suited for high-power microwave applications where traditional microwave components are shorted and remote generation is advantageous. Moreover, the plasma densities ($n_e \leq 10^{25}~$m$^{-3}$)\cite{ostrovskaya1974laser} and dimensions (size $\geq 10~\mu$m)\cite{roskos2007broadband} achieved in these resonators translate to resonance frequencies as high as a few THz, making these MTMs potentially suitable for applications in the elusive THz gap.

The first step towards understanding the scattering properties of these near-ellipsoidal resonators was to find a relationship between their resonance frequency and intrinsic plasma properties. An exact solution for the scattering fields from an ellipsoid, similar to the solution derived by G. Mie for the scattering fields from a sphere \cite{mie1908}, does not exist. However, for sub-wavelength particles, an electrostatic approximation can be used to obtain an approximate solution.  Through such an approach, the polarizability of a sub-wavelength ellipsoid in a field parallel to one of its principal axes was found to be \cite{bohren1998}
\begin{equation}
	\alpha = v\dfrac{\epsilon-1}{1+L(\epsilon-1)},
\label{eq:alpha}
\end{equation}
where $v$ is the volume of the ellipsoid, $L$ is a geometrical factor dependent on the eccentricity and orientation of the ellipsoid, and $\epsilon$ is the dielectric constant of the plasma ellipsoid, which can be represented by a Drude model, i.e.,
\begin{gather}
\label{eq:eps}
	\epsilon = 1 - \frac{\omega_p^2}{\omega^2 + i \gamma\omega}; \quad \omega_p = \sqrt{\dfrac{n_e q^2}{m_e\epsilon_0}},
\end{gather}
where $\omega$ is the incident radiation frequency, $q$ is the electron charge, $m_e$ is the electron mass, and $\epsilon_0$ is the free space permittivity. The absorption and scattering cross sections of such ellipsoids can then be written in terms of the polarizability as
\begin{equation}
\label{eq:cross}
	C_{\text{abs}} = k \text{Im}\{\alpha\}; \quad C_{\text{sca}} = \dfrac{k^4}{6\pi}|\alpha|^2,
\end{equation}
where $k$ is the wave number \cite{bohren1998}. These cross sections quantify the amount of electromagnetic radiation that is absorbed/scattered by the ellipsoids as a function of frequency. From Eq. (\ref{eq:cross}), it is possible to see that there will be a resonance (i.e. a surface mode will be excited) in both cross sections at the frequency that makes the denominator of $\alpha$ vanish, i.e., when
\begin{equation}
\label{eq:omega}
	\omega_0 = -\dfrac{i}{2}\left(\gamma + \sqrt{\gamma^2 - 4L\omega_p^2}\right)
\end{equation}
or when $\omega_0 = \omega_p\sqrt{L}$ for a collision-less ($\gamma=0$) plasma. Equation (\ref{eq:omega}) highlights the tunable nature of the described gaseous plasma resonators. The dependence of $\omega_0$ on $\omega_p$ and $\gamma$, confirms its dependence on the energy invested by the ionizing source and the pressure/composition of the gas, respectively. The geometrical factor, $L$, is independent of the plasma properties, but for the prolate-spheroid shaped plasma generated by a laser focused to a circular beam waist, it could take on two different values depending on whether the electric field is incident parallel or perpendicular to the major axis of the spheroid. The geometrical factors for the two different cases are defined as
\begin{gather}
\label{eq:L_par}
	L_\parallel = \dfrac{1-e^2}{e^2}\left(-1 + \dfrac{1}{2e}\ln\dfrac{1+e}{1-e}\right); \\
\label{eq:L_per}
	L_\perp = \dfrac{1}{4e^3}\left(2e + (e^2-1)\ln\dfrac{1+e}{1-e}\right),
\end{gather}
where $e$ is the eccentricity of the prolate spheroid \cite{bohren1998}. By substituting Eqs. (\ref{eq:eps}) and (\ref{eq:L_par}) [or Eq. (\ref{eq:L_per})] into Eq. (\ref{eq:omega}), the resonance frequency for a plasma spheroid with major axis oriented parallel [or perpendicular] to the incident field was found to be dependent on $n_e$, $\gamma$, and $e$. It is important to recognize that for any finite collisionality case, Eq. (\ref{eq:omega}) will have a complex solution. However, provided Im($\omega_0$) $\ll$ Re($\omega_0$), Re($\omega_0$) provides a good approximation for the resonance frequency of the resonator. 

\begin{figure}[t]
	\centering
  	\includegraphics[scale=1]{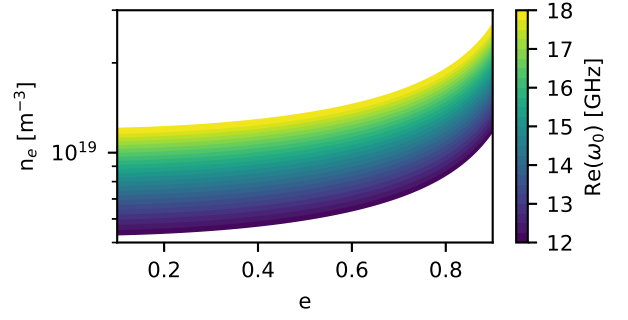}
 	\caption{\label{fig:ne_vs_e} Range of values of $n_e$ and $e$ leading to resonance in the Ku band for the case where the incident electric field is parallel to the major axis of the resonator and $\gamma = 0$.}
\end{figure}

In this letter, the conditions necessary for resonance in the Ku band (12-18~GHz) of the microwave spectrum were explored for a LIP with major axis oriented parallel to the incident electric field. Only LIPs with dimensions $\leq$~4mm, which are commonly reported in literature, were considered in order to satisfy the long wavelength criteria required by both Eq.~(\ref{eq:omega}) and the effective-medium limit. Figure~\ref{fig:ne_vs_e} shows contour plots of the range of values of $n_e$ and $e$ which would lead to resonances inside this frequency range for a collisionless plasma. As shown in the figure, resonances could be achieved in the Ku band for resonators with densities ranging from $n_e \approx 5\times 10^{18}~$m$^{-3}$ to $n_e \approx 3\times 10^{19}~$m$^{-3}$ assuming an eccentricity range of 0.1 $<$ e $<$ 0.9. It is important to recognize that for the more realistic case of a collisional plasma, an increase in $\gamma$ will lead to a redshift in these values of Re($\omega_0$), provided Im($\omega_0$) $\ll$ Re($\omega_0$). 

\begin{figure}[b!]
	\centering
  	\includegraphics[width=8.4cm]{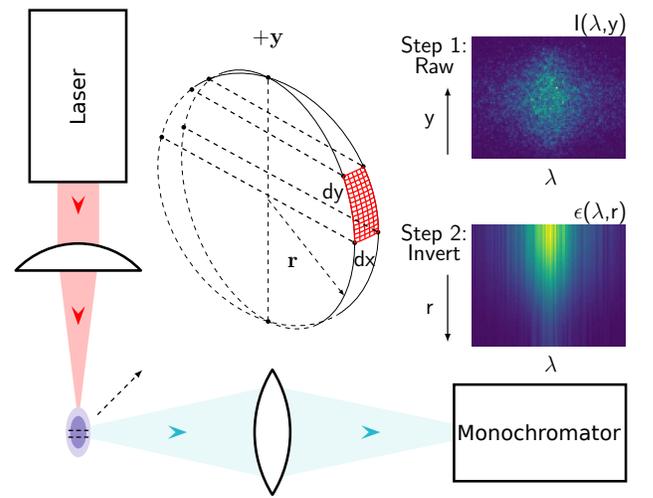}
 	\caption{\label{fig:spectroscopy_setup} Schematic of experimental setup for stark broadening measurements, together with sample images of the LIP's chord-integrated intensity and Abel-inverted emissivity.}
\end{figure}

Optical emission measurements were carried out to determine if values in this range of $n_e$ and $e$ could be achieved for a low collisionaliy LIP. A schematic of the experimental setup used for these measurements can be found on Fig.~\ref{fig:spectroscopy_setup}. As shown in the figure, a collimated IR beam generated by a Q-switched Nd:YAG laser (emission wavelength: 1064~nm, output energy: $\sim$200~mJ/p, repetition rate: 10~Hz, pulse duration: $\sim$10~ns, beam diameter: $\sim$10~mm) was focused through a plano-convex lens (diameter: 30~mm, focal length: 40~mm) to produce a plasma spheroid. The plasma was generated in a 93\%N$_2$/7\%H$_2$ gas mixture at a relatively low pressure of 20 Torr to minimize electron-neutral collision losses. The time evolution of $n_e$ was determined by measuring the stark broadening of the LIP's H$\alpha$ emission line and the eccentricity was determined from gated images of its broadband emission. The LIP was imaged at the entrance slit (width = 60~$\mu$m) of a Jobin-Yvon U1000 double monochromator where a dual holographic 1800~groove/mm grating dispersed the light, which was then imaged onto a Princeton Instruments PI-MAX intensified CCD camera. A wavelength calibration factor of 0.0568~\r{A}/pixel and height calibration factor of 0.0307~mm/pixel were determined by placing a mercury spectral lamp at the focus of the plasma-generating lens and focusing on two closely spaced Hg lines. Two 1~mm black strips taped onto the mercury lamp 1~mm apart were used to calibrate the vertical spatial resolution. The instrument broadening was determined using a HeNe laser and was used to establish a baseline for deconvolution from broadening due to the plasma. The camera was triggered simultaneously with the laser Q-switch and gated to a 150~ns exposure so as to capture the time evolution of the plasma emission.

In order to determine the radial profiles of the H$\alpha$ line emission from the chordwise integrated emission intensity, the inverse Abel transform was utilized, i.e.,
\begin{gather}
\label{eq:emission}
	\epsilon(r) = -\dfrac{1}{\pi} \int_r^R \dfrac{dI(y)}{dy}\dfrac{dy}{\sqrt{y^2-r^2}},
\end{gather}
where $I(y)$ is the experimentally measured, chord-integrated intensity, $R$ is the radius at which the measured intensity reaches the background level, and $\epsilon(r)$ is the calculated radial emissivity. To compute Eq.~(\ref{eq:emission}), the Nestor-Olsen method was used, i.e.,
\begin{gather}
	\epsilon_k(r) = -\dfrac{2}{\Delta y\pi}\sum_{n-k}^{N-1}I(y_n)B_{k,n},
\end{gather}
where the integers $k$ and $n$ are the position indices of the radial and vertical intensities, and $\Delta y$ is the distance between adjacent experimental data points. For details of the calculation of the weights $B_{k,n}$, see Ref. \cite{nestor1960numerical}. The Abel transform is highly sensitive to the symmetry of the measured integral data. Accordingly, the measured intensity was symmetrized about the centroid of the Stark broadened H$\alpha$ emission line. Each vertical slice of the symmetrized data was fit with a Gaussian curve and inverted, as shown in Fig.~\ref{fig:spectroscopy_setup}. After the emissivity profile $\epsilon(\lambda,r)$ was reconstituted from the inverted slices, a Voigt profile was fit to each radial slice. The Lorentzian component of the Voigt fit is due to the Stark broadening, from which the plasma density was calculated using an empirical correlation for the Stark broadening of hydrogen \cite{ehrich1980experimental}. 

\begin{figure}[t!]
	\centering
  	\includegraphics[scale=1]{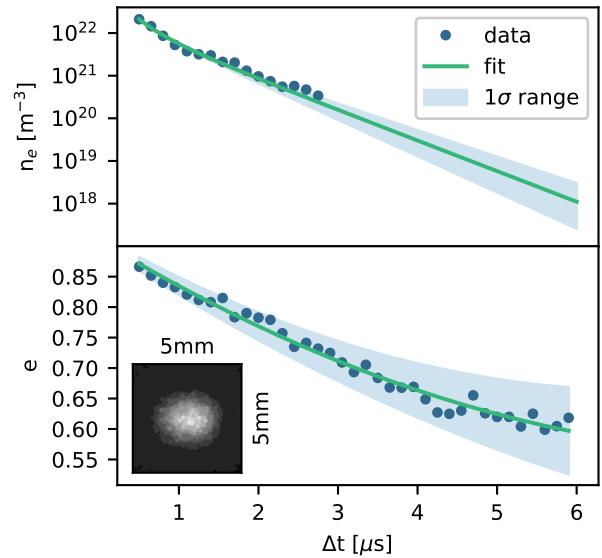}
 	\caption{\label{fig:spectroscopy_results} Time evolution of $n_e$ and $e$ as measured by optical diagnostic experiments, together with a sample broadband emission image of the LIP taken at $\Delta$t = 5~$\mu$s.}
\end{figure}

A plot of the time evolution of the radially-averaged $n_e$ resulting from these measurements can be found on Fig.~\ref{fig:spectroscopy_results}. Due to the pulsed nature of this laser, the $n_e$ in this plasma peaked at early times and decayed through recombination/diffusion as a function of time \cite{ostrovskaya1974laser}. For this reason, the data is presented as a function of  $\Delta$t, which was the delay time between the laser Q-switch trigger and the time the measurement was taken. Data is only presented for the first $\sim$2.5~$\mu$s due to a decreased number of counts in H$\alpha$ emission at later times, which is attributed to the relatively low gas pressure set to minimize losses in subsequent microwave scattering experiments. In an effort to determine the time at which $n_e$ reached values that would lead to resonance in the Ku band, the data was fitted and extrapolated using a recombination/diffusion model, i.e.,
\begin{gather}
	\dfrac{dn_e}{dt} = -\alpha n_e^2 - \dfrac{n_e}{\tau},
\end{gather}
where $\alpha$ is the recombination coefficient and $\tau$ is the characteristic diffusion time scale. Using a non-linear least squares fit method, an optimal fit was found for $\alpha = 9.235\times10^{-17}~m^3/s$ and $\tau = 0.606~\mu s$. The resulting fit is plotted on Fig.~\ref{fig:spectroscopy_results} together with a one standard deviation ($\sigma$) interval envelope. As seen in the figure, an extrapolation of this fit to later times shows that the radially-averaged $n_e$ reaches values that would lead to resonance in the Ku band [see Fig.~\ref{fig:ne_vs_e}] for an approximate time range of 4.1-5.0~$\mu$s (1$\sigma$ range: 3.6-5.6~$\mu$s).

The time evolution of the eccentricity was measured through gated images of emission from the LIP captured by bypassing the monochromator and focusing the broadband image of the plasma directly on the intensified CCD camera. Given that this image was primarily generated by light emitted through recombination radiation and excitation of neutral species throughout the plasma, the images provided a good estimate for the geometry and eccentricity of the resonators. The resulting time evolution of $e$ measured from these 150~ns exposure images can be found on Fig.~\ref{fig:spectroscopy_results}, together with a sample image taken at t = 5~$\mu$s. The figure shows that at times where $n_e$ reaches values that would lead to resonance in the Ku band, the spheroid eccentricity is in the range of 0.62-0.66 (1$\sigma$ range: 0.54-0.72). This further reduces the range of $n_e$ values that could lead to resonance in the Ku band and in turn the approximate time range at which these resonances are expected to 4.4-4.9~$\mu$s (1$\sigma$ range: 3.9-5.5~$\mu$s). The sample image shows that at these times, the plasma size is $\sim$3~mm, which is below $\lambda/4$ for all Ku band wavelengths, satisfying the long wavelength limit required by both Eq.~(\ref{eq:omega}) and the effective-medium limit. 

\begin{figure}[t!]
	\centering
  	\includegraphics[scale=1]{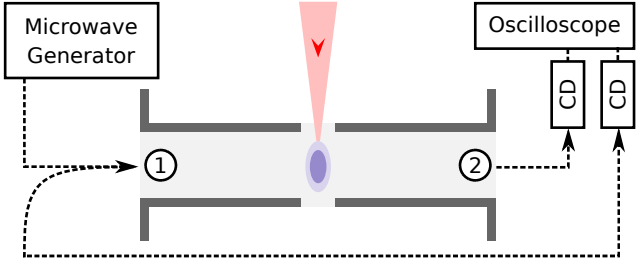}
 	\caption{\label{fig:microwave_setup} Schematic of experimental setup for Ku band microwave scattering measurements.}
\end{figure}

Microwave scattering experiments were carried out to confirm the existence of the described resonance in the LIP at the times suggested by the optical diagnostics. A schematic of the measurement setup can be found on Fig.~\ref{fig:microwave_setup}. The same experimental conditions used for the optical emission measurements were used to produce a plasma spheroid with major axis oriented parallel to the incident electric field inside a WR62 waveguide. The incident microwave signal was generated by an HP 83732A signal generator and reflection/transmission signals were detected by Krytar 303SK crystal detectors (CDs) and recorded through an oscilloscope.

\begin{figure}[b!]
	\centering
  	\includegraphics[scale=1]{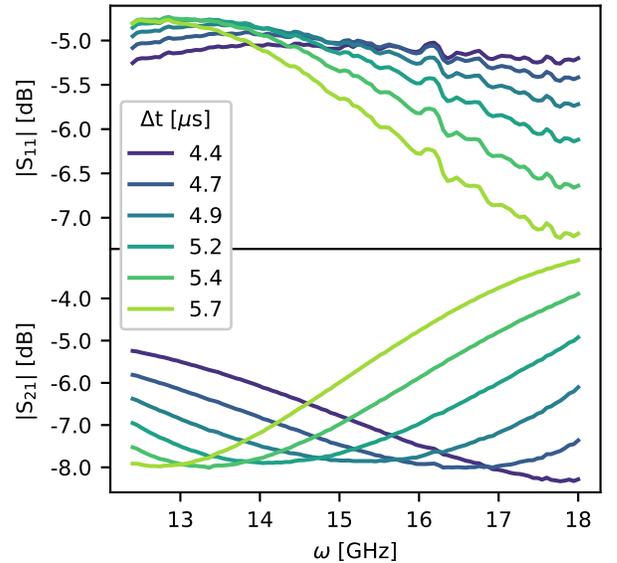}
 	\caption{\label{fig:spectra} Experimental reflection and transmission spectra of a WR62 waveguide embedded with a single LIP element.}
\end{figure}

Plots of reflection ($|S_{11}|$) and transmission ($|S_{21}|$) coefficient vs. $\omega$ obtained from this two-port waveguide experiment can be found on Fig.~\ref{fig:spectra} for several values of $\Delta$t. The plots show that resonant behavior for the given laser and gas conditions is observed in the Ku band at laser Q-switch delay times in the range of 4.4-5.7~$\mu$s, which is mostly within the standard error of the optical emission measurements. Discrepancies in $\Delta$t were expected given the approximate nature of Eq.~(\ref{eq:omega}) and the finite collisionality/size of the LIP. The plots show a clear trend of decreasing $\omega_0$ (i.e. frequency of least transmission) with increasing $\Delta$t (decreasing $n_e$), following the trend theorized in Eq.~(\ref{eq:omega}). The resonant behavior is supported by clear extinction spectra with minimum power ratios of $\sim$15\% and corresponding reflection spectra with peak power ratios of $\sim$30\%, which points to absorption values of $\sim$55\%. The distribution of energy between reflection and absorption is likely dependent on $\gamma$, with decreasing values of $\gamma$ leading to increased scattering/reflection. This dependency will be studied further in future publications. 

The results presented thus far confirm the feasibility of the described gaseous plasmonic resonators. Such resonators could be used to extrapolate the functionality of solid-state Mie-based MTMs \cite{zhao2009mie} to extreme conditions where energy fluxes exceed the damage threshold of solid-state materials and remote generation is advantageous. The most simple application of these resonators would be as a meta-atom in a tunable microwave or THz metasurface (MTS) mirror or absorber. Such a MTS could be generated by focusing a high-energy laser pulse through a micro-lens array and into a gas. This and other examples of all-plasma MTMs will be explored in future publications. 

In summary, this letter addressed the potential use of LIPs as tunable resonators for MTM applications. An analytical model was presented describing the scattering resonance of these near-ellipsoidal plasmas and its dependence on their eccentricity and intrinsic plasma properties. This dependence was confirmed through Ku band transmission experiments of a waveguide with an embedded single plasma element and through optical diagnostics of the LIP. The described resonator has the potential to be used as the building block in a new class of MTMs with fully three-dimensional structural flexibility in the microwave and THz regime of the EM spectrum. 

\bibliography{mybib}

\end{document}